\documentclass[12pt]{article}

\usepackage{latexsym}

\newcommand{\ft}[2]{{\textstyle\frac{#1}{#2}}} \newsavebox{\uuunit}
\sbox{\uuunit}
    {\setlength{\unitlength}{0.825em}
     \begin{picture}(0.6,0.7)
        \thinlines
        \put(0,0){\line(1,0){0.5}}
        \put(0.15,0){\line(0,1){0.7}}
        \put(0.35,0){\line(0,1){0.8}}
        \multiput(0.3,0.8)(-0.04,-0.02){12}{\rule{0.5pt}{0.5pt}}
        \end {picture}}

\def\lra{\leftrightarrow}

\csname @addtoreset\endcsname{equation}{section}

\def\ts{\textstyle}
\def\sst{\scriptscriptstyle}

\def\del{\partial}
\def\ds{\not \!\del}
\def\cDs{\not\!\!{\cal D}}

\newcommand{\w}[1]{\\[0.#1cm]}

\def\be{\begin{equation}}
\def\ee{\end{equation}\\[-.75cm] }
\def\ba{\begin{array}}
\def\ea{\end{array}}
\def\bea{\begin{eqnarray}}
\def\eea{\end{eqnarray}\\[-.75cm] }
\def\bd{\begin{document}}
\def\ed{\end{document}}

\let\la=\label

\let\bm=\bibitem
\def\nn{\nonumber}
\def\qq{\quad\quad}

\def\ft#1#2{{\textstyle{{\scriptstyle #1}\over {\scriptstyle #2}}}}
\def\fft#1#2{{#1 \over #2}}

\def\sst#1{{\scriptscriptstyle #1}}
\def\oneone{\rlap 1\mkern4mu{\rm l}}

\newcommand{\eq}[1]{Eq.~(\ref{#1})}
\newcommand{\sect}[1]{Sec.~(\ref{#1})}

\newcommand{\eqs}[2]{Eqs.~(\ref{#1})-(\ref{#2})}

\def\Hat#1{\widehat{#1}}

\def\a{\alpha}
\def\b{\beta}
\def\c{\gamma}
\def\C{\Gamma}
\def\d{\delta}
\def\D{\Delta}
\def\e{\epsilon}
\def\vare{\varepsilon}
\def\eb{{\bar\epsilon}}
\def\F{\Phi}
\def\vf{\varphi}
\def\k{\kappa}
\def\l{\lambda}
\def\L{\Lambda}
\def\m{\mu}
\def\n{\nu}
\def\r{\rho}
\def\s{\sigma}
\def\S{\Sigma}
\def\th{\theta}
\def\Th{\Theta}
\def\o{\omega}
\def\O{\Omega}

\def\cD{{\cal D}}
\def\cF{{\cal F}}
\def\cM{{\cal M}}
\def\cR{{\cal R}}

\def\ket#1{\left|#1\right>}
\def\sc#1#2{[\,#1\,,\,#2\,]_{\star}}

\def\fa{
\begin{table}[htb]
\la{tbl:tensorMultiplet}
\begin{center}
\begin{tabular}{||c|c|c|c|c||}
\hline
Field & Type & Restrictions & {\rm USp(4)} & w\\
\hline
&&&&\\
$B_{\m\n}$& boson&real antisymmetric tensor gauge field&1&0\\
&&&&\\
$\psi^i$& fermion &$\c_7\psi^i = -\psi^i$&4& ${\ts{5\over
2}}$\\
&&&&\\
$\phi^{ij}$&boson&$\phi^{ij} = -\phi^{ji} \ \ \ \
\Omega_{ij}\phi^{ij} = 0$&5&2\\
&&&&\\
\hline
\end{tabular}
\caption{\footnotesize {\it Fields of the $(2,0)$ tensor multiplet.}
We have indicated the various algebraic restrictions on the fields,
their USp(4) representations assignments and the Weyl weights $w$.}
\end{center}
\end{table} }

\def\fb{
\begin{table}[h!]
\la{tbl:fieldsWeyl}
\begin{center}
\begin{tabular}{||c|c|c|c|c||}
\hline
Field & Type & Restrictions & {\rm USp(4)} & w\\
\hline
&&&&\\
$e_\m{}^a$&boson&sechsbein&1&-1\\
&&&&\\
$\psi_\m^i$&fermion&$\c_7\psi_\m^i = +\psi_\m^i$&4& -${\ts{1\over2}}$\\
&&&&\\
$V_\m^{ij}$&boson&$V_\m^{ij} = V_\m^{ji}$&10&0\\
&&&&\\
$T_{abc}^{ij}$&boson&$T_{abc}^{ij} = -T_{abc}^{ji}$ &5&1\\
&&&&\\
&&$\Omega_{ij}T_{abc}^{ij} = 0 $&&\\ &&&&\\ &&$T^{ij}_{abc} =
-{\ts{1\over 6}}\e_{abcdef}T^{ij}_{def}$&&\\
&&&&\\
$\chi^{ij}_k$&fermion&$\chi^{ij}_k = -\chi^{ji}_k$ &16& ${\ts{3\over 2}}$\\
&&&&\\
&&$\Omega_{ij}\chi^{ij}_k = 0$ &&\\
&&&&\\
&&$\chi^{ij}_i= 0 $&&\\
&&&&\\
&&$\c_7 \chi^{ij}_k = +\chi^{ij}_k$&&\\
&&&&\\
$D^{ij,k\ell}$&boson&$D^{ij,k\ell}=-D^{ji,k\ell} = -D^{ij,\ell k}$ &14&2\\
&&&&\\
&&$D^{ij,k\ell} = D^{k\ell,ij} $&&\\ &&&&\\ &&$\Omega_{ij}D^{ij,k\ell} =
\Omega_{k\ell}D^{ij,k\ell} = 0 $&&\\
&&&&\\
&&$\Omega_{ik}\Omega_{j\ell}D^{ij,k\ell } = 0$&&\\
&&&&\\
\hline
\end{tabular}
\caption{{\footnotesize {\it Fields of $(2,0)$ conformal supergravity:}
We have indicated the various algebraic restrictions on the fields,
their USp(4) representation assignments, and the Weyl weights $w$. A
field $\phi$ of weight $w$ transforms under dilatations as
$\delta_D\phi = w\L_D \phi.$}}
\end{center}
\end{table}
}

\begin{document}

\begin{titlepage}

\begin{flushright}

KUL-TF-99/13\\

UG/4-99\\

CTP TAMU-10/99\\

hep-th/9904085

\end{flushright}

\vspace{.5cm}

\begin{center}

\baselineskip=16pt

{\LARGE  \bf (2,0) Tensor Multiplets and Conformal Supergravity in D=6}
\end{center}

\bigskip\bigskip

\centerline{{\large \bf Eric Bergshoeff $^1$, Ergin Sezgin $^2$ and
Antoine Van Proeyen $^3$}}

\vspace{2cm}

{\footnotesize $^1$ Institute for Theoretical Physics, Nijenborgh 4,
9747 AG Groningen,The Netherlands \\

$^2$ Center for Theoretical Physics, Texas A\&M University, College
Station, Texas 77843--4242, U.S.A.\\

$^3$ Instituut voor Theoretische Fysica, Katholieke Universiteit
Leuven, Celestijnenlaan 200D, B-3001 Leuven, Belgium}

\vfill

\begin{center}

{\bf Abstract}

\end{center}

{We construct the supercurrent multiplet that contains the
energy-momentum tensor of the $(2,0)$ tensor multiplet. By coupling
this multiplet of currents to the fields of conformal supergravity, we
first construct the linearized superconformal transformations rules of
the $(2,0)$ Weyl multiplet. Next, we construct the full non-linear
transformation rules by gauging the superconformal algebra
$OSp(8^*|4)$. We then use this result to construct the full equations
of motion of the tensor multiplet in a conformal supergravity
background. Coupling $N+5$ copies of the tensor multiplet to conformal
supergravity and imposing a geometrical constraint on the scalar fields
which fixes the conformal symmetry, we obtain the coupling of $(2,0)$
Poincar\'e supergravity to $N$ tensor multiplets in which the physical
scalars parametrize the coset $SO(N,5)/(SO(N)\times SO(5))$.}

\vspace{2mm}

\vfill

\hrule width 3.cm

{\footnotesize
\noindent \\
$^2$ Supported in part by the U.S. National Science
Foundation, under grant PHY-9722090. \\
\noindent $^3$ Onderzoeksdirecteur FWO, Belgium.}

\end{titlepage}


\section{Introduction}


There is an increasing evidence for the fact that $M$-theory on anti de
Sitter (AdS) backgrounds can be described by a conformal field theory
at the boundary of AdS, at least in a suitable limit. The low energy
limits involved in the bulk of AdS are the gauged supergravity theories
in various dimensions and the boundary field theories are certain
globally supersymmetric field theories appropriate to the branes
involved.\\

In verifying the AdS/CFT correspondence, the boundary values of the
bulk fields naturally arise. In accordance with the fact that AdS
supersymmetry in a given dimension acts as the conformal
supersymmetry at the boundary of AdS, the boundary values of the bulk
fields are in one-to-one correspondence with the fields of conformal
supergravity defined at the boundary. Thus, it is natural to formulate
the boundary field theory in a conformal supergravity background.
Integration over the boundary (matter) fields should then yield an
effective action involving the conformal supergravity fields, which is
to be compared with the bulk supergravity effective action.\\

This approach was followed in \cite{liu1}, where the coupling of
$N=4,D=4$ super Yang-Mills to $N=4,D=4$ conformal supergravity
\cite{roo1} was studied, as the boundary field theory associated with
gauged supergravity in $AdS_5$. In this spirit, we wish to construct
the coupling of the $(2,0)$ tensor multiplet to $(2,0)$ conformal
supergravity at the six dimensional boundary of $AdS_7$. This result is
expected to provide a convenient framework in studying the $AdS_7/CFT_6$
correspondence. Since the $(2,0)$ tensor multiplet contains a chiral
$2$-form, it is natural to study its field equations rather than an
action from which they may be derivable but which may require the
introduction of additional fields. Thus, we shall primarily study the
covariant field equations, although we shall briefly discuss an action
from which all but the self-duality condition follows, provided that the
self-duality equation is imposed after the variation.\\

The conformal supergravity fields form an off-shell multiplet. We can
treat them as background fields, in which case we need not impose their
equations of motion. However, coupling of conformal supergravity to
$N+5$ copies of the $(2,0)$ tensor multiplet, we can constrain the
scalars fields coming from these tensor multiplets so that they become
a representative of the coset ${SO(N,5)\over SO(N)\times SO(5)}$. As we
will show, this leads to a conformal interpretation of the
$(2,0)$ Poincar\'e supergravity coupled to $N$ tensor multiplets
constructed previously in \cite{romans,ric}.\\

The organization of this paper is as follows. In Sec. 2, we derive the
multiplet of supercurrents and the linearized Weyl multiplet. In Sec.
3, we construct the full $(2,0)$ conformal supergravity theory. As we
did before for the $(1,0)$ case \cite{ber1}, we follow the methods
developed first for $N=1$ in 4 dimensions \cite{superconformN1}. They
are based on gauging the conformal superalgebra \cite{SU221}, which, in
our case, is $OSp(8^*|4)$. As is typical in this method, one then has
to impose constraints on some of the curvatures. In Sec. 4, we find the
complete equations of motions for a single tensor multiplet in a
conformal supergravity background. In Sec. 5, we consider $N+5$ copies
of the tensor multiplet in conformal supergravity background and show
that a geometrical constraint on the scalars leads to the equations of
motion of Poincar\'e supergravity coupled to $N$ tensor multiplets. Further
comments on our results and open problems are collected in the
Conclusions. Our notations and conventions are presented in Appendix A
and the truncation of our results to the $(1,0)$ case \cite{ber1} are
described in Appendix B, as we used that correspondence to obtain our
present results. A superspace description of the tensor, current and
Weyl multiplets is given in Appendix C.


\section{ The $(2,0)$ Supercurrent Multiplet}\la{tm}

In this section we will construct the $(2,0)$ supercurrent multiplet.
Using the invariance of the bilinear couplings between the
currents and the corresponding fields we will derive the {\it linearized}
transformation rules of the (2,0) conformal supergravity multiplet.
In the next section we will extend this to the nonlinear case.\\

Our starting point is the $(2,0)$ tensor multiplet in $D=6$ Minkowski
spacetime describing $8+8$ degrees of freedom\footnote{
In this section the tensor multiplet will only play an auxiliary role
as a means to construct the supercurrent multiplet. Later, in
section~\ref{tm2}, (2,0) tensor multiplet will be introduced  as matter multiplets
to be coupled to the conformal supergravity theory constructed in section~\ref{ss:Weylm}.}.
This is the only
on-shell $(2,0)$ matter multiplet in $D=6$. Its field components are
given in Table~\ref{tbl:tensorMultiplet}. It contains a $2$-form
potential $B_{\m\n}$ whose self-dual field strength is defined by

\be
{H}_{\m\n\rho} = 3\del_{[\m}B_{\n\rho]}\ ,
\qq
H_{\m\n\r} = \ft1{3!} \e_{\m\n\r\s\l\tau}\, H^{\s\l\tau}\equiv
\tilde H_{\mu \nu \rho }\ .
\label{defHdual}
\ee

It also contains $5$ real scalars $\phi^{ij}\, (i=1,...,4)$ which
transforms as a $5$-plet of $USp(4)$ and a symplectic Majorana-Weyl
spinor $\psi^i$. The basic properties of these fields are tabulated in
Table 1.\\

\fa

The complete rigid superconformal transformations have been given in
\cite{cla1} where the tensor multiplet was studied as the $M5$-brane
worldvolume supermultiplet. For our present purposes we only need the
linearized rigid $Q$-supersymmetry transformation rules

\bea
\d B_{\m\n} &=& - {\bar\e}\c_{\m\n}\psi \ ,
\nn\w2
\d\psi^i &=& {\ts{1\over 48}} H^{+}_{\m\n\r} \c^{\m\n\r} \e^i
+{\ts{1\over 4}}\not\! \del \phi^{ij} \e_j\ ,
\la{tr}
\w2
\d \phi^{ij} &=& -4 {\bar\e}^{[i}\psi^{j]} -\O^{ij}\bar \e \psi \ .
\nn
\eea

The self-dual part of the curvature $H^+_{abc}$ transforms as

\be
\d H_{abc}^{+} = -\ts{1\over 2}{\bar \e}\ds \c_{abc}\psi\ .
\ee

The supersymmetry transformations close provided that the following
linearized field equations are satisfied

\be
H^- = \ds \psi^i = \del^\m\del_\m \phi^{ij} = 0\ .
\la{fe}
\ee

To construct the current multiplet, we start from the Noether currents
of the $(2,0)$ tensor multiplet. These Noether currents are: the
energy-momentum tensor $\theta_{\m\n}$, the supersymmetry currents
$J_{\m i}$ and the ${\rm USp(4)}$ currents $v_\m^{ij}$. We will use the
improved currents that satisfy the following equations:

\bea
&& \del^\m\theta_{\m\n} = 0\ ,\hskip 1.5truecm \th_{\m\n}=\th_{\n\m}\ ,
\hskip 1.5truecm \th^\m{}_\m =0\ ,
\nn
\w2
&& \del^\m J_{\m i}= 0\ , \hskip 1.2truecm \c^\m J_{\m i} = 0\ ,
\nn
\w2
&& \del^\m v_{\m }^{ij} = 0\, .
\la{conditions}
\eea

These symmetry properties determine the currents up to constants, which
we have determined by requiring the closure of the rigid linearized
supersymmetry transformations. We thus find the currents

\bea
\la{currents}
\th_{\m\n} &=&  H^+_{\m ab}H^+_{\n}{}^{ab}
+ 8 {\bar \psi}\c_{(\m}\del_{\n)}\psi + \del_\m
\phi^{ij}\del_\n\phi_{ij}-\ft1{10} \eta_{\m\n} \left
(\del\phi\right )^2 -{\ts{1\over 5}}
\del_\m\del_\n \phi^2\ ,
\nn\w2
J_{\m i} &=& -\ft23 H^+_{\r\s\tau} \c^{\r\s\tau}\c_\m\psi_i
+8\phi_{ij}\stackrel{\lra}{\del_\m}\psi^j
+\ft85\c_{\m\l}\del^\l \left (\phi_{ij}\psi^j\right )\, ,
\w2
v_\m^{ij} &=& -2\phi_k{}^{(i}{\del}_\m\phi^{j)k} +
8{\bar\psi}^i\c_\m\psi^j\ ,
\nn
\eea

where $\phi^2 \equiv \phi^{ij}\phi_{ij}$. The currents are conserved
provided that the fields satisfy the free field equations \eq{fe}. When
we apply supersymmetry transformations \eq{tr} on the currents
\eq{currents}, always using the field equations \eq{fe}, we find a full
supermultiplet of operators bilinear in the fields:\footnote{The same
operators have been given in \cite{how1} using the superfield approach.}

\bea
\d \th_{\m\n} &=&\ft14 {\bar\e}\c_{\r(\m}\del^\r J_{\n)}\ ,
\nn
\w2
\d J_\m^i &=&  \c^\n\theta_{\m\n}\e^i
-\ft18\left(\c^\r\c_{\m\n} - {\ts{3\over 5}}
\c_{\m\n}\c^\r\right)\del^\n v_\r^{ij} \e_j
\nn
\w2
&& -\ft1{4}\left(\c ^{abc}\c _{\m\n }
+\ft15 \c _{\m\n}\c ^{abc}\right) \del^\n t_{abc}^{ij}\e _j\ ,
\nn\w2
\d v_\m^{ij} &=& -\ft12{\bar\e}^{(i}J_\m^{j)}
+\ft{15}8{\bar \e}_k \c_{\m\n}\del^\n\l^{k(i,j)}\ ,
\la{deltatheta}
\w2
\d t_{abc}^{ij} &=& -\ft1{24} {\bar\e}^{[i}\c^\rho \c_{abc} J_\rho^{j]}
+\ft5{32} {\bar\e}^k \not\!\del \c_{abc}\l_k^{ij} - ({\rm trace})\, ,
\nn
\w2
\d\l^{ij}_k &=&- \ft1{15} \c^{abc}t^{ij}_{abc}\e_k - \ft4{15}\c^\m v_{\m k}{}^{[i}
\e^{j]} -4\not\!\del d^{ij}_{kl}\e^l - ({\rm trace})\ ,
\nn\w2
\d d^{ij,kl} &=&-\ft{1}8 {\bar\e}^{[i}\l^{j],kl}
+ (ij \lra kl) -({\rm trace})\ ,
\nn
\eea

where we have introduced the operators  $\l ^{ij}_k$, $t_{abc}^{ij}$and
$d^{ij,kl}$ defined by

\bea
\la{operators}
t_{abc}^{ij} &=& \ft23 H^+_{abc}\phi^{ij}-\ft43{\bar\psi}^i\c_{abc}\psi^j
-\ft13\Omega^{ij} {\bar\psi}\c_{abc}\psi\ ,
\nn
\w2
\l^{ij}_k &=& -\ft{32}{15}\phi^{ij}\psi_k
-\ft{128}{75}\d^{[i}_k\phi ^{j]\ell }\psi _\ell
+\ft{32}{75} \O^{ij}\phi_{k\ell }\psi ^\ell \ ,
\w2
d^{ij}_{k\ell } &=& -\ft1{15}\phi^{ij}\phi_{k\ell } +\ft1{75} \d^{[i}_{[k}
\d^{j]}_{\ell ]}\phi ^2 -\ft1{300}\O^{ij}\O_{k\ell}\phi ^2\ .
\nn
\eea

Note that the operator $t_{abc}^{ij}$ is {\it self-dual}. To prove the
transformation rules we have used several $USp(4)$ Schouten identities
such as

\bea
&& 2\bar \e ^{[i}\phi ^{j]k}\psi _k+2\bar \e ^k\phi _k{}^{[i}\psi ^{j]}
  +\phi ^{ij}\bar \e \psi - ({\rm trace})=0\ ,
\nn
\w2
&&\c^{abc}\e_k\,{\bar\psi}^i \c_{abc}\psi^j
-2\c^{abc}\e^{[j}\,{\bar\psi}^{i]}\c_{abc}\psi_k -({\rm traces}) =0\ .
\la{Schouten}
\eea

The currents \eq{currents} and the operators \eq{operators} constitute
the multiplet of currents. Taking into account the conditions
\eq{conditions} this supermultiplet contains $128 + 128$ degrees of
freedom. Using this current multiplet, the linearized Weyl multiplet is
derived by introducing the Noether coupling

\be
\int d^6x \left[ h_{\m \n }\theta ^{\m \n }+ \bar \psi ^\m J_\m
+ V_\m ^{ij}v^\m _{ij} + T_{abc}^{ij}t^{abc}_{ij}
+\bar \chi ^{ij}_k\l ^k_{ij} +D^{ij,kl} d_{ij,kl}\right]\ .
\la{invariant}
\ee

The independent conformal supergravity fields which have been introduced
above are

\be
h_{\m\n}\ ,\ \psi_\m^i\ ,\ V_\m^{ij}\ ,\ T_{abc}^{ij}\ ,\
\chi^i_{jk}\ ,\ D^{ij, kl}\ .
\la{fc}
\ee

Several properties of these gauge and matter fields are summarized in
Table~\ref{tbl:fieldsWeyl}. \\

\fb

Demanding the invariance of the action \eq{invariant} under rigid
supersymmetry, we find the following linearized Weyl multiplet
transformation rules:

\bea
\d h_{\m\n} &=& \eb\c_{(\m}\psi_{\n)}\ ,
\nn\w2
\d \psi_\m^i &=& -\ft14 (\del_\r h_{\m\s})\c^{\r\s}\e^i
+\ft12 V_\m^{ij}\e_j +\ft1{24}T^{ij}_{abc} \c^{abc}\c_\m \e_j\ ,
\nn
\w2
\d V_\m^{ij} &=&
\ft14\eb^{(i}\left(\c^{\r\s}\c_\m-\ft35\c_\m\c^{\r\s}\right)\psi_{\r\s}^{j)}
-\ft4{15}\eb_k\c_\m\chi^{(i,j)k}\ ,
\nn
\w2
\d T_{abc}^{ij} &=&
\ft18\eb^{[i}\left(\c^{\m\n}\c_{abc}+\ft15\c_{abc}\c^{\m\n}\right)
\psi_{\m\n}^{j]} -\ft1{15}\eb^k\c_{abc}\chi_k^{ij} -({\rm trace})\ ,
\nn
\w2
\d\chi^{ij}_k &=& \ft5{32}
\left(\del_\m T^{ij}_{abc}\right)\c^{abc}\c^\m \e_k
-\ft{15}8\c^{\m\n} V_{\m\n k}{}^{[i}\e^{j]}
-\ft14 D^{ij}_{k\ell}\e^\ell -({\rm traces})\ ,
\nn
\w2
\d D^{ij,k\ell} &=& -2 \eb^{[i} \ds \chi^{j],k\ell}
-2 \eb^{[k} \ds \chi^{\ell],ij} -({\rm trace})\ .
\eea

where the supersymmetry parameter is constant, and we have defined

\bea
\psi_{\m\n} &=& \del_\m\psi_\n -\del_\n \psi_\m\ ,
\nn\w2
V_{\m\n}^{ij} &=& \del_\m V_\n^{ij} -\del_\n V_\m^{ij}\ .
\eea

In the next section, we generalize the above transformation rules to
obtain the full local superconformal transformation rules.


\section{The (2,0) Conformal Supergravity Theory }\la{ss:Weylm}

In this section, we will construct the nonlinear (2,0) conformal
supergravity theory. Our starting point is the linearized conformal
multiplet constructed in the previous section. This multiplet contains
both gauge fields and matter fields (not to be confused with
the tensor multiplet matter fields that will be coupled in section~\ref{tm2}).
Due to the presence of the matter fields, the nonlinearization cannot be
understood as a straightforward gauging of an underlying superconformal
algebra. To include the matter fields, one must follow a
6-step procedure that has been explained in detail in \cite{ber1}. In the same
reference, this 6-step procedure has been applied to construct the
(1,0) conformal supergravity theory.  Here we apply the same procedure
to construct the (2,0) theory.\\

The (2,0) conformal supergravity is based on the superconformal algebra $OSp(8^*|4)$ whose
generators are labeled

\be
T_A = P_a\ , Q_{\a i}\ , U_{ij}\ , M_{ab}\ , K_a\ , S_{\a i}\ , D\ ,
\ee

where $a,b, \cdots$ are Lorentz indices, $\a$ is a chiral spinor index
and $i,j = 1, \cdots 4$ are $USp(4)$ indices. $M_{ab}$ and $P_a$ are
the Poincar\'e generators, $K_a$ is the special conformal
transformation, $D$ the dilatation, $Q_{\a i}$ and $S_{\a i}$ are the
supersymmetry and special supersymmetry generators, respectively, which
are symplectic Majorana-Weyl spinors, 16 real components in total.
Finally, $U^{ij} = U^{ji}$ are the ${\rm USp(4)}$ generators. For more
details on the $OSp(8^*|4)$ algebra and the rigid superconformal
transformations, see \cite{cla1}. \\

The gauge fields corresponding to the above generators are

\be
e_\m{}^a\ ,\psi_\m^i\ ,\ V_\m^{ij}\ ,  \o_\m{}^{ab}\  ,
\ f_\m{}^a\ ,\phi_\m^i\ ,\ b_\m\ .
\ee

However, in the realization (Weyl multiplet) which will gauge the
algebra, these fields are not all independent. The independent fields
are given in \eq{fc}, where the first three are the gauge fields
corresponding to the generators $P_a\ , Q_{\a i}$ and $U_{ij}$. It is
understood that the linearized gravitational field $h_{\m\n}$ has been
replaced by the sechsbein $e_\m{}^a$. The last three are matter fields
needed for the realization of the superconformal algebra.
The remaining gauge fields are either dependent $(\o_\m{}^{ab}\ ,
f_\m{}^a\ ,\phi_\m^i)$ (see below), or can be shifted away
($b_\m$) using $K_a$ invariance.\\

The $(2,0)$ Weyl multiplet describes $128 + 128$ off-shell degrees of
freedom. We first present the result and next explain our notation and
give our definitions. The bosonic transformations of the independent
gauge fields are given by general coordinate transformations and

\bea
\d e_\m{}^a &=& -\L_D e_\m{}^a - \L^{ab}e_{\m b}\ ,
\nn
\w2
\d \psi_\m^i &=& -{\ts{1\over 2}}\L_D \psi_\m^i
+{\ts{1\over 2}}\L^i{}_j\psi_\m^j -{\ts{1\over 4}}\L^{ab}\c_{ab}\psi_\m^i\ ,
\nn
\w2
\d V_\m^{ij} &=& \del_\m\L^{ij} +\L^{(i}{}_kV_\m^{j)k}\, ,\nn
\w2
\d b_\m &=& \del _\m \L_D-2e_\m ^a\L_{K\,a }\ .
\la{BosTransfWeyl}
\eea

where $\L_D, \,\L_{K\,a },\,\L^{ab}$ and $\L^{ij}$ are the parameters
of dilatation, special conformal, Lorentz and USp(4) transformations,
respectively. The transformation properties of the matter fields,
$T,\chi $ and $D$ under dilatations, Lorentz and USp(4) transformations
 follow from the rules (\ref{BosTransfWeyl}) and from
Table~\ref{tbl:fieldsWeyl}. All matter fields are inert under the
special conformal transformations $K$. \\

Following \cite{superconformN1,ber1} we impose the following curvature
constraints
\footnote{Note that in contradistinction to the (1,0) case,
discussed in \cite{ber1}, the matter fields $\chi^i_{jk}$ and $D$ are
absent in the constraints. See also Appendix~\ref{app:truncation}.}

\bea
R_{\m\n}{}^a (P) &=& 0\, ,
\nn
\w2
\la{constraints}
R_{\m\n}{}^{ab}(M) e^\n{}_b + {\ts{1\over 4}} T_{\m bc}^{ij}
T^{abc}_{ij} &=& 0\, ,
\w2
\c^\m R_{\m\n}^i(Q) &=& 0\, .\nn
\eea

The above matter-modified curvatures are defined by

\bea
R_{\m\n}{}^a(P) &=& 2\del_{[\m}e_{\n]}^a + 2b_{[\m}e_{\n]}^a
+{\underline { 2\omega_{[\m}{}^{ab}e_{\n]b}}} -{\ts{1\over 2}}{\bar
\psi}_\m\c^a\psi_\n\ ,
\nn\w2
R_{\m\n}{}^{ab}(M) &=&2\del_{[\m}\omega_{\n]}{}^{ab} + 2
\omega_{[\m}{}^{ac}\omega_{\n]c}{}^b
-{\underline {8f_{[\m}{}^{[a}e_{\n]}{}^{b]}}}
+{\bar\psi}_{[\m}\c^{ab}\phi_{\n]}
\nn
\w2
&&+{\bar\psi}_{[\m}\c^{[a}R_{\n]}{}^{b]}(Q) +{\ts{1\over 2}} {\bar
\psi}_{[\m}\c_{\n]}R^{ab}(Q) +{\ts{1\over
2}}{\bar\psi}_{\m,i}\c_c\psi_{\n,j} T^{abc,ij}\, ,
\la{curvatures}
\w2
R_{\m\n}^i(Q) &=& \left ( 2 \del_{[\m}\psi_{\n]}^i +b_{[\m}\psi_{\n]}^i
+ {\ts{1\over 2}} \omega_{[\m}{}^{ab}
\c_{ab}\psi_{\n]}^i - V_{[\m}^i{}_j \psi_{\n]}^j\right )\nn
\w2
&&+ {\underline { 2\c_{[\m}\phi_{\n]}^i}} +{\ts{1\over 12}}
T^{ij}_{abc}\c^{abc}\c_{[\m}\psi_{\n]j}\ .
\nn
\eea

The underlined terms indicate all terms of the form gauge field
$\times$ sechsbein. Since the sechsbein is invertible the corresponding
gauge fields can be solved for from the constraints \eq{constraints}.
Explicitly, the constraints \eq{constraints} enable us to solve for the
gauge fields ($\omega_\m{}^{ab}\ ,f_\m{}^a\ ,\phi_\m^i)$ as follows:

\bea
\o_\m{}^{ab} &=& 2e^{\n[a}\del_{[\m}e_{\n]}{}^{b]}
- e^{\rho [a}e^{b]\sigma}e_\m{}^c\del_\rho e_{\s c}
\nn
\w2
&& + 2 e_\m{}^{[a} b^{b]} +{\ts{1\over 2}}{\bar\psi}_\m\c^{[a}\psi^{b]}
+ {\ts{1\over 4}}{\bar\psi}^a\c_\m\psi^b\ ,
\nn
\w2
f_\m{}^a &=& -{\ts{1\over 8}}R^{\prime }_\m{}^a(M)
+{\ts{1\over 80}}e_\m{}^a R^\prime (M)
+{\ts{1\over 32}}T^{ij}_{\m cd}T_{ij}^{acd}\ ,
\w2
\phi_\m^i &=&-{\ts{1\over 16}}\left (\c^{ab}\c_\m
-{\ts{3\over 5}}\c_\m\c^{ab}\right )R^{\prime }_{ab}{}^i(Q)\ .
\nn
\eea

The notation $R^\prime$ indicates that in the corresponding curvature
the underlined term in \eq{curvatures} has been omitted, and
$R'_\mu {}^a=e_b^\nu R'_{\mu \nu }{}^{ba}$.\\

We next give the full non-linear $Q$ and $S$-transformations of the
$(2,0)$ Weyl multiplet:

\bea
\d e_\m{}^a &=&{\ts{1\over 2}}\bar\e\c^a\psi_\m\ ,
\nn
\w2
\d b_\m &=& -\ft12\bar\e\phi _\m +\ft12\bar \eta \psi_\m
\nn\w2
\d \psi_\m^i &=& \cD_\m\e^i + {\ts{1\over 24}}
T^{ij}_{abc}\c^{abc}\c_\m\e_j + \c_\m\eta^i\ ,
\nn
\w2
\d V_\m^{ij} &=& -4{\bar\e}^{(i} \phi_\m^{j)} -
{\ts{4\over 15}}{\bar\e}_k\c_\m\chi^{(i,j)k} -4
{\bar\eta}^{(i}\psi_\m^{j)}\ ,
\nn
\w2
\la{N=4}
\d T_{abc}^{ij}&=& {\ts{1\over 8}}{\bar\e}^{[i}
\c^{de}\c_{abc}{ R}^{j]}_{de}(Q)
- {\ts{1\over 15}} {\bar\e}^k\c_{abc}\chi^{ij}_k - ({\rm trace})\ ,
\w2
\d \chi_k^{ij} &=&{\ts{5\over 32}} \left (
\cD_\m T^{ij}_{abc}\right )\c^{abc}\c^\m\e_k -{\ts{15\over 16}}
\c^{\m\n} R_{\m\n k}{}^{[i}(V) \e^{j]} -{\ts{1\over 4}}D^{ij}_{kl}\e^l
\nn
\w2
&&+{\ts{5\over 8}} T^{ij}_{abc}\c^{abc}\eta_k -({\rm traces})\ ,
\nn
\w2
\d D^{ij,kl} &=& - 2
{\bar\e}^{[i}\cDs \chi^{j],kl} + 4 {\bar\eta}^{[i}\chi^{j],kl}
+ (ij\lra kl ) -({\rm trace})\ .
\nn
\eea

We have used here the following definitions. The covariant derivatives
and $USp(4)$ curvature in (\ref{N=4}) are:

\bea
\cD_\m \e^i &=& \del_\m\e^i +{\ts{1\over 2}}b_\m\e^i
+{\ts{1\over 4}}\o_\m{}^{ab}\c_{ab}\e^i -{\ts{1\over 2}}V_{\m}{}^i{}_j\e^j\ ,
\nn
\w2
R_{\m\n}{}^{ij}(V) &=& 2\del_{[\m}V_{\n]}{}^{ij} +
V_{[\m}{}^{k(i}V_{\n]}{}^{j)}{}_k +8{\bar\psi}_{[\m}{}^{(i}
\phi_{\n]}{}^{j)}
+{\ts{8\over 15}}{\bar\psi}_{[\m,k}\c_{\n]}\chi^{(i,j)k}\ .
\la{de}
\eea

The supercovariant derivatives $\cD_\m$ of matter fields are defined as
the ordinary derivative $\del_\m$ plus a covariantization term which is
always given by minus all the transformation rules of the matter field
with the parameter replaced by the corresponding gauge field. For
example the supercovariant derivative of $T$ is given by

\bea
\cD_\m T_{abc}^{ij} &=& \del_\m T_{abc}^{ij} +3\omega_{\m [a}{}^d
T_{bc]d}^{ij} - b_\m T_{abc}^{ij} +V_\m^{[i}{}_k T_{abc}^{j]k}
\w2
&& -{\ts{1\over 8}}{\bar\psi}_\m^{[i}
\c^{de}\c_{abc}{ R}^{j]}_{de}(Q)
+ {\ts{1\over 15}} {\bar\psi}_\m^k\c_{abc}\chi^{ij}_k - ({\rm trace})\,
.
\nn
\eea


\section{The $(2,0)$ Tensor Multiplet in the Conformal
Supergravity Background}
\la{tm2}


In this section we couple a (2,0) tensor multiplet to conformal
supergravity. Our starting point will be the linearized transformation
rules of the tensor multiplet. The nonlinear rules can then be obtained
by imposing the superconformal algebra via an iterative Noether procedure.
This procedure has been described in detail for the $(1,0)$
case in \cite{ber1}. The same procedure can be applied here.
As an alternative, we will derive the same result by the
requirement that the (2,0) nonlinear tensor multiplet should reproduce,
upon truncation the (1,0) nonlinear tensor multiplet of \cite{ber1}.
The details of this truncation are explained in Appendix~\ref{app:truncation}.\\

Our starting point is the
the linearized equations of motion and the supersymmetry transformations
of the $(2,0)$ tensor multiplet given in section~\ref{tm}.
Applying the truncation procedure described in Appendix~\ref{app:truncation}, we find that
the full nonlinear $Q$ and $S$-transformations of the $(2,0)$ tensor
multiplet are given by:

\bea
\d B_{\m\n} &=& - {\bar\e}\c_{\m\n}\psi +{\bar\e}^i
\c_{[\m}\psi_{\n]}^j\phi_{ij}\ ,
\nn
\w2
\d \psi^i &=& {\ts{1\over 48}} { H}^{+}_{\m\n\r}\c^{\m\n\r} \e^i
+{\ts{1\over 4}}\not\!\! \cD \phi^{ij}
\e_j - \phi^{ij}\eta_j\ ,
\nn
\w2
\d \phi^{ij} &=& -4
{\bar\e}^{[i}\psi^{j]} -({\rm trace})\ .
\la{N=4T}
\eea

The curvature $H_{\m\n\r}$ is defined by

\be
H_{\m\n\rho} = 3\del_{[\m}B_{\n\rho]}
+3{\bar\psi}_{[\m}\c_{\n\rho]}\psi
-{\ts{3\over2}}{\bar\psi}^i_{[\m}\c_\n\psi^j_{\rho]} \phi_{ij}\ .
\ee

It satisfies the Bianchi identity

\be
\cD_{[a}H_{bcd]} - {\ts{3\over 2}} {\bar\psi}\c_{[ab} {R}_{cd]}(Q)=0\ .
\ee

The self-dual part of the curvature ${ H}^+_{abc}$ transforms as

\be
\delta { H}_{abc}^{+} = -{\ts{1\over 2}}{\bar \e}
\not\!\! \cD\c_{abc}\psi -3 {\bar\eta}\c_{abc}\psi\ .
\ee

Furthermore, we find that the field equations of the $(2,0)$ tensor
multiplet are given by\footnote{
Note that, in contradistinction to the (1,0) case the first field equation
can {\it not} be used to solve for the matter field $T$ in terms of $H^-$
(the scalar $\phi$ is not a singlet under USp(4)). Therefore, the (2,0)
Weyl multiplet has {\it no} alternative formulation containing an antisymmetric
tensor gauge field like the (1,0) Weyl multiplet (see \cite{ber1}).}

\bea
{\cal F}^-_{abc} &:=& { H}^-_{abc} - {\ts{1\over 2}}
\phi_{ij}T_{abc}^{ij} = 0\ ,
\nn
\w2
\C^i &:=& \not\!\! \cD\psi^i
-{\ts{1\over 15}}\phi^{kl}}\chi^i_{kl} -{\ts{1\over 12}
T^{ij}_{abc}\c^{abc}\psi_j = 0\ ,
\la{fe2}
\w2
{\cal C}_{ij} &:=& \cD^a \cD_a \phi_{ij} -{\ts{1\over
15}}D_{ij}^{kl}\phi_{kl} +{\ts{1\over 3}}{ H}^+_{abc} T_{ij}^{abc} +
{\ts{16\over 15 }}{\bar \chi}_{ij}^k\psi_k = 0\ ,
\nn
\eea

where

\bea
\cD_\m \psi^i &=& \left (\del_\m - {\ts{5\over 2}} b_\m +
{\ts{1\over 4}}\o_\m{}^{ab}\c_{ab}\right )\psi^i -{\ts{1\over
2}}V_\m^i{}_j\psi^j\nn
\w2
&&-{\ts{1\over 48}}H^+_{abc}\c^{abc}\psi_\m^i - {\ts{1\over 4}}\left (
\not\!\! \cD\phi^{ij}\right )
\psi_{\m j} + \phi^{ij}\phi_{\m j}\ ,
\w2
\cD_\m \phi^{ij} &=& \left (\del_\m -2b_\m\right )
\phi^{ij} + V_\m{}^{[i}{}_k \phi^{j]k} + 4 \left (
{\bar\psi}_\m^{[i} \psi^{j]} - {\rm trace}\right )\ .
\nn
\eea

To determine the supercovariant d'Alembertian of the scalars we first
calculate the transformation properties of $\cD_a\phi^{ij}$:

\bea
\d \cD_a\phi^{ij} &=& 3\L_D  \cD_a\phi^{ij}
+ \L^{[i}{}_k \cD_a\phi^{j]k} -\L_a{}^b \cD_b\phi^{ij}
+4\L_{Ka}\phi^{ij}
\nn
\w2
&&-4 {\bar\e}^{[i}\cD_a \psi^{j]} +{\ts {2\over 15}}\left(
{\bar\e}_l\c_a\chi^{(i,k)l}\phi^j {}_k - i\lra j\right)
\w2
&&+{\ts{1\over 6}}{\bar\e}_k\c_a \c^{bcd}T_{bcd}^{k[i}
\psi^{j]}
-4 {\bar\eta}^{[i}\c_a \psi^{j]} - ({\rm trace})\ .
\nn
\eea

{}From this we derive that

\bea
\la{box}
\cD^a\cD_a \phi^{ij} &=& \del^a\cD_a \phi^{ij}
-3b^a \cD_a\phi^{ij} + V_a^{[i}{}_k \cD^a\phi^{j]k} +\o_a{}^{ab}
\cD_b\phi^{ij} -4 f_a{}^a\phi^{ij}\nn
\w2
&&+4 {\bar \psi}_a^{[i}\cD^a \psi^{j]} -{\ts {2\over 15}}\left (
{\bar\psi}^a_l\c_a\chi^{(i,k)l}\phi^j {}_k - i\lra j\right )
\w2
&&-{\ts{1\over 6}}{\bar\psi}^a_k\c_a \c^{bcd}T^{k[i}_{bcd}
\psi^{j]}
+4 {\bar\phi}_a^{[i}\c^a \psi^{j]} - ({\rm trace})\ .\nn
\eea

Note the occurrence of the Riemann curvature scalar in the equation
of motion for the scalar fields through the term $f_a{}^a\phi_{ij}$.
This contains as well graviton as gravitino terms of the supergravity
action, the latter through $f_a^ a=-\ft1{20}R'(M)+\ldots
=\ft1{160}\bar \psi _\mu \gamma ^{\mu \nu \rho }R'_{\nu \rho
}(Q)+\ldots $. Gravitino kinetic terms appear also in the equation of
motion for tensor multiplet fermions through the term
$\not\!\!{\cal D}\psi ^i=\phi^{ij}\gamma ^\mu \phi_{\m j}+\ldots =
\ft1{10}\phi^{ij}\gamma ^{\mu \nu }R'_{\mu \nu j}(Q)+\ldots$.\\

While it is possible to compute the Green's functions for the tensor
multiplet fields in presence of the conformal supergravity background
by starting from the equations of motion, it
would be convenient to perform such calculations by starting from an
action. To construct a manifestly Lorentz invariant action
requires the introduction of an auxiliary scalar field \cite{pst}. It
has been shown in \cite{cla1} that this can straightforwardly be
implemented in a rigid conformal theory.
We expect that this can be extended for the local superconformal
case.\\

An
alternative approach is to relax the chirality condition on the
$2$-form potential and to write an action which is not invariant, but
whose variation is proportional to $ {\cal F}^-_{abc}$. It
gives the correct equations of motion provided that
the self-duality condition $ {\cal F}^-_{abc}=0$ is imposed after the action is varied
\cite{e,c,ric,p}. Such an action takes the form

\be
S=\int d^6 x \left(- \ft16 H^{abc} \cF^-_{abc} -4{\bar\psi}^i\C_i
- {\bar\psi}_\m^i \c^\m \C^j \phi_{ij}
+\ft14 \phi^{ij} {\cal C}_{ij}\right) \ .
\ee


\section{Poincar\'e Supergravity Coupled to  $N$ (2,0) Tensor
Multiplets from the Superconformal Theory}\la{ps}

In this section we construct the matter couplings to $(2,0)$
supergravity by using the superconformal tensor calculus and by
imposing the $SO(N,5)$ symmetry. We hereby closely follow the
procedure of coupling $N=4$, $d=4$ vector multiplets as in
\cite{deRooWag}. This procedure was first introduced in \cite{conformforPoin}
and has been applied to obtain matter couplings in 4 dimensions for $N=1$
\cite{N1YMmsg}, $N=2$ \cite{dWLVP} and $N=4$ \cite{ber2}.
The basic idea is that there is a close relation between
matter-coupled Poincar\'{e} and conformal supergravity theories.
Starting for (a slight generalization of)
the matter-coupled conformal supergravity theory
constructed in the previous section, we simply gauge fix the conformal
scale and S-supersymmetry transformations to reproduce
existing results on D=6 matter-coupled Poincar\'{e} supergravity
\cite{romans,ric}. For a review of this technique, see for example
\cite{review}.\\

We begin by introducing $(N+5)$ copies of the $(2,0)$ tensor multiplets
with fields $B_{\m\n}^I\ , \psi^{iI}\ , L_I^{ij}$ where $I=1,...,N+5$
labels the vector representation of $SO(N,5)$. We have denoted the
scalars by $L_I^{ij}$ because they will shortly be constrained. The
constraint will be solved in terms of independent scalar fields which
will again be denoted by $\phi$.\\

The superconformal transformation rules now read

\bea
\d B^I_{\m\n} &=& - {\bar\e}\c_{\m\n}\psi^I +{\bar\e}^i
\c_{[\m}\psi_{\n]}^j L^I_{ij}\ ,
\nn
\w2
\d \psi^{iI} &=& {\ts{1\over 48}} { H}^{I+}_{abc}\c^{abc} \e^i
+{\ts{1\over 4}}\not\!\! \cD L^{Iij}\e_j - L^{Iij}\eta_j\ ,
\nn
\w2
\d L^{Iij} &=& -4 {\bar\e}^{[i}\psi^{j]I} - \O^{ij}{\bar\e}\psi^I  \ .
\la{5N}
\eea

Since the tensor multiplet fields occur linearly in the full field
equations \eq{fe2}, the latter generalize to the case of $N+5$ tensor
multiplets as

\bea
&& { H}^{I-}_{abc} - {\ts{1\over 2}} L^I_{ij}T_{abc}^{ij} = 0\ ,
\nn
\w2
&& \not\!\! \cD\psi_I^i -{\ts{1\over 15}}L_I^{kl}}\chi^i_{kl}
-{\ts{1\over 12} T^{ij}_{abc}\c^{abc}\psi_{Ij} = 0\ ,
\la{fe3}
\w2
&& \cD^a \cD_a L^I_{ij} -{\ts{1\over 15}}D_{ij}^{kl}L^I_{kl}
+{\ts{1\over 3}}{ H}^{I+}_{abc} T_{ij}^{abc} +{\ts{16\over 15 }}{\bar
\chi}_{ij}^k\psi^I_k = 0\ .
\nn
\eea

Note that the index $I$ is a global $SO(N+5)$ index and consequently
the derivatives of $L^I_{ij}$ and $\psi^I_i$ occurring in \eq{5N} and
\eq{fe3} are as defined earlier for $\phi_{ij}$ and $\psi$ without any
new connection terms to rotate the index $I$.

To obtain the Poincar\'e supergravity coupled to $N$ copies of the
$(2,0)$ tensor multiplet, we impose the geometrical constraint

\bea
\eta^{IJ} L_I^{ij} L_{Jk\ell} &=&
 -\d^{[i}_{[k}\d^{j]}_{\ell ]}+\ft14\O^{ij}\O_{k\ell}\
\nn\w2
&\equiv& \eta^{ij}{}_{k\ell}
\la{c}
\eea

where $\eta_{IJ}$ is a symmetric invariant tensor of $SO(N,5)$ with
signature $(-----++\cdots +)$. The raising and lowering of the
$SO(N,5)$ indices will always be done with the metric $\eta_{IJ}$. The
condition \eq{c}, together with the fact that $L_I^{ij}$ are defined up
to local $USp(4)$ transformations, reduces the number of independent
scalars to $(N+5)\times 5 - 15-10 =5N$, which is the dimension of the
coset ${SO(N,5)\over SO(N)\times SO(5)}$. It is convenient to introduce
an $(N+5)\times N$ matrix $L_I^r\ (r=1,...,N)$ which together with
$L_I^{ij}$ form an $(N+5)\times (N+5)$ matrix $L_I^A$ satisfying the
condition

\be
\eta^{IJ} L_I{}^A L_{JB} = \eta^A{}_B\ ,
\ee

where $A=(ij,r)$ and the $\eta^A{}_B$ is the constant metric with
components: $\eta^{ij}{}_{k\ell}\ , \eta^r{}_s=\d^r{}_s$ and
$\eta^{ij}{}_r=\eta^r{}_{ij}=0$. \\

The constraint \eq{c} is invariant under $S$--supersymmetry. However,
varying it under $Q$--supersymmetry gives the constraint

\be
L_I^{ij} \psi^I_k = 0\ .
\la{c2}
\ee

This constraint is easily solved as

\be
\psi^i_I = L_I^r \psi^{ri}\ ,
\la{psi}
\ee

where $\psi^{ri}\,(r=1,...,N)$ are the independent fermionic fields.\\

Next, we vary the traceless part of \eq{c2} to obtain the constraint

\be
L_I^{ij}\cD_\m L^I_{k\ell} =
-8{\bar\psi}^{I[i}\c_\m\psi_{I[k}\d_{\ell]}^{j]} \ ,
\la{v}
\ee

and, making use of \eq{Schouten}, the equation of motion for the
$2$-form potential

\be
H_{abc}^{I+} L_I^{ij} =-2{\bar\psi}^i_I \c_{abc}\psi^{jI}\ .
\la{he}
\ee

Requiring that the trace part of the constraint \eq{c2} is invariant
under the combined $Q$ and $S$--transformations, and using
\eqs{psi}{he} and performing Fierz re-arrangement, we determine the
$S$--supersymmetry parameter:

\be
\eta_i=
-\ft12({\bar\psi}_i^r\c^a\psi^k_r)\,\c_a\e_k
-\ft1{72} ({\bar\psi}_i^r\c^{abc}\psi^k_r)\,\c_{abc}\e_k
-\ft1{36}({\bar\psi}^k_r \c^{abc}\psi_k^r)\,\c_{abc}\e_i \ .
\la{eta}
\ee

Next, we observe that $V_\m^{ij}$ can be solved from \eq{v} as

\be
V_{\mu i}{}^j = 2 L^I_{ik} {\cal D'}_\m L_I^{jk}
-8{\bar\psi}_i^I\c_\m\psi_I^j\ ,
\ee

where ${\cal D'}_\m$ is the supercovariant derivative without the
$V_\m^{ij}$ term. The Weyl multiplet fields $T^{ij}_{abc},\chi^i_{jk}$
and $D^{ij}_{k\ell}$ are also readily solved from
\eq{fe3}. For example,

\be
T^{ij}_{abc}= -2 H^{I-}_{abc} L_I^{ij}\ .
\la{t}
\ee

Using $K$--symmetry, we can also set

\be
b_\m=0\ .
\ee

The independent fields we are left with are those of the combined
$(2,0)$ Poincar\'e supergravity plus $N$ tensor multiplet system,
namely:

\be
e_\m{}^a\ , \psi_\m^i\ , B_{\m\n}^I\ , \psi_i^r\ , L_I^A\ .
\ee

The Poincar\'e supersymmetry transformations of these fields can be
found from \eq{N=4} and \eq{N=4T} by using the solutions for the Weyl
multiplet fields and the compensating $S$--supersymmetry transformation
\eq{eta}. We thus find

\bea
\d e_\m{}^a &=&{\ts{1\over 2}}\bar\e\c^a\psi_\m\ ,
\nn
\w2
\d \psi_\m^i &=& \cD_\m\e^i - {\ts{1\over 12}}
L_I^{ij}H^{I-}_{\r\s\tau}\c^{\r\s\tau}\c_\m\e_j + \c_\m\eta^i\ ,
\nn
\w2
\d B^I_{\m\n} &=& -  L^I_r {\bar\e}^i\c_{\m\n}\psi_i^r
+L^I_{ij}{\bar\e}^i \c_{[\m}\psi_{\n]}^j \ ,
\la{final}
\w2
\d \psi^{ir} &=& {\ts{1\over 48}} L_I^r { H}^{I+}_{\m\n\r}\c^{\m\n\r} \e^i
+{\ts{1\over 4}} V_\a^{r,ij} \cDs \phi^\a \e_j
-\d\phi^\a A_\a^{rs} \psi_s^i\ ,
\nn
\w2
\d\phi^\a &=& 4 V^\a_{r,ij}\,{\bar\e}^i\psi^{jr}\ .
\nn
\eea

where $\phi^\a\ (\a=1,...,5N)$ are the scalar fields parametrizing the
coset ${SO(N,5)\over SO(N)\times SO(5)}$ and $\eta^i$ is given in
\eq{eta}. The vielbein $V_\a^{r,ij}$, the $SO(N)$ connection
$A_\a^{rs}$ and the $USp(4)$ connection $A_\a^{ij}$ on this coset are
defined as

\bea
V_\a^{r,ij} &=& L^{Ir}\del_\a L_I^{ij}\ ,
\nn\w2
A_\a^{rs} &=& L^{Ir}\del_\a L_I^s\ ,
\nn\w2
A_{\a i}{}^j &=& 2 L^I_{ik} \del_\a L_I^{jk}\ .
\eea

Further definitions are as follows:

\bea
H^I_{\m\n\rho} &=& 3\del_{[\m}B^I_{\n\rho]}
+3{\bar\psi}_{[\m}\c_{\n\rho]}\psi^r L^I_r
-{\ts{3\over2}}{\bar\psi}^i_{[\m}\c_\n\psi^j_{\rho]} L^I_{ij}\ ,
\nn
\w2
\cD_\m \phi^\a &=& \del_\m \phi^\a -4V^\a_{r,ij}\,{\bar\psi}_\m^i\psi^{jr}\ ,
\la{defs2}
\w2
\cD_\m \e^i &=& \del_\m\e^i +{\ts{1\over 4}}\o_\m{}^{ab}\c_{ab}\e^i
-{\ts{1\over 2}}V_{\m}{}^i{}_j\e^j\ ,
\nn
\eea

where the composite connection $V_\m^{ij}$ is given by

\be
V_\m^{ij} = \cD_\m \phi^\a A_\a^{ij} -8{\bar\psi}^i_r\c_\m\psi^{jr}\ .
\ee

Comparing the result \eq{final} with that of \cite{ric}, we find that
all the structures are in agreement except the last term in $\d
\psi^{ir}$, which is missing in \cite{ric}.\\

The self-duality condition \eq{he} serves as the full field equation
for the $2$-form potential $B_{\m\n}^I$. The remaining field equations
follow from the closure of the algebra \eq{final}. The resulting field
equations can be found in \cite{romans,ric}.

Summarizing, in this section we have shown that the (2,0) matter-coupled
Poincar\'{e} theory of \cite{romans, ric} can be reproduced by fixing the
conformal gauges in the (2,0) matter-coupled conformal supergravity
constructed in this paper.

\section{Conclusions}


In this paper we have constructed the local conformal supersymmetry
rules for $(2,0)$ supergravity in 6 dimensions. That includes the
transformation rules for the Weyl multiplet \eq{N=4}, which is the
gauge multiplet of the $OSp(8^*|4)$ superconformal algebra and the
transformation laws \eq{N=4T} of the tensor multiplet. The latter has
field equations given by \eq{fe2}.
 These results can be viewed as the quadratic approximation to
the coupling of the full $M5$ brane theory to conformal supergravity in
a physical gauge. It would be interesting to obtain the full coupling
of the $M5$-brane to the $(2,0)$ conformal supergravity.\\

Taking $N+5$ copies of the tensor multiplets and imposing the
constraints described in section~\ref{ps}, reproduces
earlier results on  Poincar\'e supergravity theory
coupled to $N$ tensor multiplets \cite{romans,ric}. The generalization
of these results to the case of $N$ coincident $M5$ branes is, of
course, a nontrivial problem. \\

We expect that the results obtained in this paper will have applications
to the study of the $AdS_7/CFT_6$ correspondence. So far, very few results
exist that deal with the calculation of the correlation functions
on the boundary of $AdS_7$ \cite{C1,C2}. Clearly, much remains to be done
to develop a better understanding of this correspondence and
the (2,0) conformal supergravity ought to play a role in this process.\\

Another open problem of interest is the construction of the higher spin
operators of the $(2,0)$ tensor multiplets \cite{rozali} and their
coupling to appropriate higher spin conformal supergravity fields. Of
special interest are the operators which correspond to massless higher
spin fields in the bulk of $AdS_7$. These arise from the product of two
doubleton representations of $OSp(8^*|4)$ \cite{g}. It is natural that
these operators couple to massless higher spin representation of this
group. A field theoretic realization of a higher spin $AdS_7$
supergravity is an interesting and challenging problem at present.

\bigskip\bigskip


\section*{Acknowledgments}


\noindent

We thank Per Sundell for useful discussions, Kostas Skenderis for a
discussion during the July 1998 Amsterdam {\it Workshop on String
Theory and Black Holes}, which triggered this investigation, and
Kor Van Hoof for indicating
corrections to a first version of this work. E.B.
thanks the institutes in Leuven and Texas A\&M, and E.S. and A.V.P.
thank the Institute for Theoretical Physics at Groningen for
hospitality. E.B. and A.V.P. thank the University of Utrecht for
hospitality.
This work was supported by the European Commission TMR
programme ERBFMRX-CT96-0045, in which E.B. is associated to Utrecht.

\bigskip


\begin{appendix} \section{Notations and Conventions}\la{app:notations}


We use the same notations as in \cite{ber1}, apart from the fact that
we now use indices from 0 to 5 with signature $(-+\cdots +)$
 rather than the Pauli convention with indices from 1 to 6 with
 signature $(+\cdots +)$. Therefore the
Levi--Civita tensor is adapted. We
replace in \cite{ber1}
\begin{equation}
i\epsilon_{abcdef} \rightarrow \epsilon_{abcdef}\ ,
\end{equation}
such that we now have
\begin{equation}
\epsilon_{012345}=1=-\epsilon^{012345}\,,\qquad
\gamma _7=\gamma^0\cdots \gamma ^5=-\gamma _0\cdots \gamma _5  \,.
\end{equation}
The essential formula is as in \cite{ber1}
\begin{equation}
  \gamma _{abc}\gamma _7=-\tilde  \gamma _{abc}\,,
\label{gamma7dual}
\end{equation}
where the dual is now defined in (\ref{defHdual}).

 We raise and lower $USp(4)$ indices
with $\Omega ^{ij}$ as:

\be
\l^i=\O^{ij}\l_j\ ,\qq \l_i=\l^j \O_{ji}\ .
\ee

When $USp(4)$ indices are omitted, northwest-southeast contraction is
understood, e.g.

\be
{\bar\l} \c^{(n)} \psi = {\bar\l}^i \c^{(n)} \psi_i\ ,
\ee

where we have used the following notation

\be
\c^{(n)} = \c^{a_1\cdots a_n} =\c^{[a_1}\c^{a_2}\cdots \c^{a_n]}\ .
\ee

The anti-symmetrizations are always with unit strength. Changing the
order of spinors in a bilinear leads to the following signs

\be
{\bar \psi}^{(1)} \c^{(n)} \chi ^{(2)} = 
t_n
\ {\bar
\chi }^{(2)} \c^{(n)} \psi^{(1)}\ ,\qquad
\left\{ \begin{array}{c}
  t_n=-1\mbox{ for }n=0,3,4 \\
  t_n=1\mbox{ for }n=1,2,5,6
\end{array}\right.
\ee

where the labels $(1)$ and $(2)$ denote any $USp(4)$ representation,
e.g. $(1)=i$ and $(2)=[jk]$.

We frequently use the following Fierz rearrangement formula:

\be
\psi_j {\bar\psi}^i= -\ft14({\bar\psi}^i\c_a\psi_j) \c^a
+\ft1{48}({\bar\psi}^i\c_{abc}\psi_j) \c^{abc}\ .
\ee

The notation ``- (trace)'' denotes terms that are proportional to
either $\O^{ij}$ or $\d^i_j$ (with ``free'' indices). We use the
notation ``-({\rm traces})'' if both invariant tensors occur. For the
convenience of the reader we give below the explicit expressions of
some trace terms:

\bea
X^{ij}-({\rm trace}) &=& X^{ij}+\ft14 \O^{ij} X^k{}_k\ ,
\nn\w2
A^{ij} X_k -({\rm traces}) &=& A^{ij}X_k +\ft45 A^{\ell[i} X_\ell
\d^{j]}_k -\ft15 \O^{ij} A_{k\ell} X^\ell\ ,
\nn\w2
S_k{}^{[i} X^{j]} -({\rm traces }) &=& S_k{}^{[i} X^{j]} -\ft15
\d^{[i}_k S^{j]\ell} X_\ell\ +\ft15 \O^{ij} S_k{}^\ell X_\ell\ .
\eea

where $X^i$ and $X^{ij}$ are arbitrary $USp(4)$ tensors, while $A^{ij}$
is an antisymmetric traceless and $S^{ij}$ a symmetric tensor.


\section{The (2,0) $\rightarrow$ (1,0) Truncation}\la{app:truncation}


Many of the formulae for the $(2,0)$ Weyl and tensor multiplet can be
obtained by considering their truncations to the $(1,0)$ case and
comparing with the results of \cite{ber1}. Following \cite{conf98} the
$(2,0)$ Weyl multiplet may also be compared with the N=4, d=4 Weyl
multiplet of \cite{ber2}. \\

We first consider the $(2,0)$ Weyl multiplet. The $(2,0)$ Weyl multiplet
(\ref{N=4}) leads to the $N=2$ Weyl multiplet of \cite{ber1} (see
eq.~(2.26)) upon making the following truncations. We write $i=1,\cdots
,4 = (i =1,2, i^\prime =1,2)$, and we put

\begin{equation}
  \Omega ^{ij}=\pmatrix{\epsilon ^{ij}&0\cr 0&\epsilon ^{i'j'}}\ .
\label{OmegaEpsilon}
\end{equation}

 The non--vanishing bosonic component
fields are given by

\bea
V_\m^{ij} &=& V_\m^{ij}\ ,
\nn
\w2
T_{abc}^{ij} &=& \e^{ij}T_{abc}\, ,\hskip 1truecm T_{abc}^{i^\prime
j^\prime} = -\e^{i^\prime j^\prime}T_{abc}\ ,
\w2
D^{ij}_{kl} &=& \e^{ij}\e_{kl}D\, ,\hskip .9truecm D^{ij}_{k^\prime
l^\prime} = -\e^{ij}\e_{k^\prime l^\prime}D\ ,
\hskip .5truecm
D^{i^\prime j^\prime }_{kl} = -\e^{i^\prime j^\prime }\e_{kl}D\ ,
\nn
\w2
D^{i^\prime j^\prime }_{k^\prime l^\prime } &=& \e^{i^\prime j^\prime }
\e_{k^\prime l^\prime} D\, ,\hskip .5truecm
D^{ij^\prime}_{k l^\prime} = - {\ts{1\over 2}}
\delta^i_k \delta^{j^\prime}_{l^\prime}D\ .
\nn
\eea

For example, the first equation above means that
$V_\m^{ij'}=0=V_\m^{i'j'}$. The non-vanishing fermionic component
fields are given by

\bea
\psi_\m^i &=& \psi_\m^i\ ,
\nn
\w2
\chi_{ij}^k &=& \e_{ij}\chi^k\ ,\hskip .5truecm
\chi^k_{i^\prime j^\prime} = -\e_{i^\prime j^\prime}\chi^k\ ,
\hskip .5truecm
\chi^{k^\prime}_{i^\prime j} = -{\ts{1\over 2}}
\delta^{k^\prime}_{i^\prime}\chi_j\ .
\eea

Thus, for example, $\psi^{i'}=0$. Finally, the non--vanishing
supersymmetry parameters are given by

\be
\e^i = \e^i\ ,\hskip 1.5truecm
\eta^i = \eta^i\ ,
\ee

which means that $\e^{i'}=0=\eta^{i'}$. In comparing the truncated
result with the $(1,0)$ Weyl multiplet of \cite{ber1} two remarks are
in order. First of all the $(1,0)$ conventional constraints of
\cite{ber1} contain extra $\chi$-- and $D$-dependent terms which do not
generalize to the $(2,0)$ case. As a consequence the dependent K and S
gauge fields, obtained after truncation, differ from those of
\cite{ber1}. In order to obtain the truncated result one should replace
the K and S gauge fields of \cite{ber1} by the following expressions

\bea
\la{redefinition}
f_\m{}^a &\rightarrow& f_\m{}^a +{\ts{1\over 240}}e_\m{}^aD\ ,
\nn
\w2
\phi_\m^i &\rightarrow& \phi_\m^i -{\ts{1\over 60}}\c_\m\chi^i \ .
\eea

Secondly, in order to remove the $\chi$--dependent term from the
supersymmetry variation of the dilatation gauge field $b_\m$ (again
this term cannot be generalized to the $(2,0)$ case) one must perform a
field-dependent $K$-transformation on the results of \cite{ber1} with
the following parameter

\be
\la{comp}
\l_{K\m} = -{\ts{1\over 60}}{\bar\e}\c_\m\chi\ .
\ee

The net effect of these manipulations is that all $\chi$-dependent
terms in the $(1,0)$ theory that cannot be extended to the $(2,0)$ case are
being removed.\\

Next we consider the (2,0) tensor multiplet.
The truncation of this multiplet to the (1,0) case treated in \cite{ber1}
is given by:

\bea
\phi^{ij} &=& \e^{ij}\sigma\, ,\hskip .5truecm
\phi^{i^\prime j^\prime} = -\e^{i^\prime j^\prime}\sigma\, ,
\hskip .5truecm \phi^{ij^\prime} = 0\, ,\nn
\w2
\psi^i &=& \psi^i\, ,\hskip 1truecm \psi^{i^\prime} = 0\, .
\eea

In order to show that the $\phi^{ij}$ field equation (see third
equation of \eq{fe2}) truncates correctly to the $(1,0)$ equation (see
eq.~(3.27) of \cite{ber1}) one has to take special care of the $\phi
D,\, {\bar\psi}_\m\chi\phi$ and ${\bar\chi}\psi$ terms. Concerning the
$D\phi$ term, starting from the $(1,0)$ case, the redefinition
\eq{redefinition} of $f_a{}^a$ leads to an extra $D\sigma$ term
which is added to the explicit $D\sigma$ term in the equation of motion
(3.27) of \cite{ber1}. As for the ${\bar\psi}_\m\chi\phi$ terms, the
redefinition of $\phi_\m$ (see \eq{redefinition}) in the
${\bar\psi}_\m{\cD}\psi$ term plus the compensating K
transformation given in \eq{comp} lead to two extra
${\bar\psi}_\m\chi\phi$ terms such that the total contribution cancels.
This is consistent with the fact that the truncation of the
${\bar\psi}_\m\chi\phi$ term in \eq{box} vanishes identically. Finally,
the redefinition of $\phi^\m$ in the ${\bar\psi}\phi_\m$ term in $
\cD^a \cD_a$ (see eq.~(3.30) of \cite{ber1}) leads to
an extra ${\bar\chi}\psi$ term which should be added to the explicit
such contribution in the $\sigma$ field equation. The total then agrees
with the $(1,0)$ truncation of our result \eq{fe2}.


\section{Tensor, Current and Weyl Multiplets in Superspace}


The $(2,0)$ tensor multiplet in flat superspace can be described by a
superfield $\phi^{ij}$ satisfying the constraint \cite{how1}

\be
D_\a^i \phi^{jk} = \Omega^{i[j}\l_\a^{k]} - ({\rm trace})\, .
\ee

In flat superspace the current multiplet \eq{deltatheta} is described
by the supercurrent \cite{how1}

\be
J^{ij,kl} = \phi^{ij}\phi^{kl} - ({\rm trace})\ ,
\ee

where the superfield $\phi^{ij}$ describes the $(2,0)$ tensor
multiplet.\\

In superspace the Weyl multiplet \eq{N=4} is described by an
anti-selfdual superfield $W_{abc}^{ij}$ in the $\bf 5$ of USp(4), whose
first component is the bosonic field $T_{abc}^{ij}$ and which satisfies
the constraint

\be
D_{\a i} W_{abc}^{jk} = \delta _i^{[j} \left (\c^{de}\c_{abc}\chi
^{k]}_{de}\right )_\a + \left (\c_{abc}\l^{jk}_i\right )_\a
 - ({\rm trace})\ ,
\ee

where $\l_i^{jk}$ is in the $\bf 16$ of $USp(4)$.

\end{appendix}

\newpage



\begin{thebibliography}{99}


\bm{liu1} H. Liu and A.A. Tseytlin, {\sl D=4 Super Yang-Mills, D=5 gauged
     supergravity and D=4 conformal supergravity},
Nucl. Phys. {\bf B533}, 88 (1998),
hep-th/9804083
     {\tt hep-th/9804083}.

\bm{roo1} M. de Roo, {\sl Matter couplings in N=4 supergravity},
      Nucl. Phys. {\bf B255} (1985) 515; \\
      M. de Roo and P. Wagemans, {\sl Gauge matter coupling in N=4
      supergravity}, Nucl. Phys. {\bf B262} (1985) 644.

\bm{romans} L.J. Romans, {\sl Self-duality for interacting fields:
      covariant field equations for six dimensional chiral supergravities},
      Nucl. Phys. {\bf B276} (1986) 71.

\bm{ric}
F.~Riccioni,
{\sl Tensor multiplets in six-dimensional (2,0) supergravity},
Phys. Lett. {\bf B422} (1998) 126,
\texttt{hep-th/9712176}.

\bm{ber1} E. Bergshoeff, E. Sezgin and A. Van Proeyen,
    {\sl Superconformal tensor calculus and matter couplings in six dimensions},
    Nucl. Phys. {\bf B264} (1986) 653.

\bm{superconformN1} M. Kaku, P.K. Townsend and P. van Nieuwenhuizen,
   {\sl Properties of conformal supergravity}, Phys. Rev. {\bf D17} (1978) 3179.

\bm{SU221} S. Ferrara, M. Kaku, P.K. Townsend and P. van Nieuwenhuizen,
    {\sl  Gauging the graded conformal group with unitary internal symmetries},
    Nucl. Phys. {\bf B129} (1977) 125.


\bm{cla1} P. Claus, R. Kallosh, A. Van Proeyen, {\sl M 5-brane and
    superconformal (0,2) tensor multiplet in 6 dimensions},
    Nucl. Phys. {\bf B518} (1998) 117, {\tt hep-th/9711161}.

\bm{how1} P.S. Howe, G. Sierra and P.K. Townsend, {\sl Supersymmetry
    in six dimensions}, Nucl.~Phys.~{\bf B221} (1983) 331.

\bm{pst}  P. Pasti, D. Sorokin and M. Tonin, {\sl Covariant action for a
    D=11 fivebrane with the chiral field}, Phys. Lett. {\bf B398} (1997) 41,
   {\tt  hep-th/9701037}.

\bm{e} E. Bergshoeff, M. de Roo and T. Ort\'{\i}n, {\sl The eleven dimensional
   fivebrane}, Phys. Lett. {\bf B386} (1996) 85, {\tt hep-th/9606118}.

\bm{c} M. Cederwall, B.E.W. Nilsson and P. Sundell, {\sl An action for
   the super-5-brane in $D=11$ supergravity},
JHEP {\bf 04}, 007 (1998), {\tt hep-th/9712059}.


\bm{p} E. Sezgin and P. Sundell, {\sl Aspects of the M5-Brane },
    {\tt hep-th/9902171}.
\bibitem{deRooWag} M. de Roo, \textsl{Matter coupling in $N=4$ supergravity},
Nucl. Phys. {\bf B255} (1985) 515;\\
P. Wagemans, \textsl{Aspects of $N=4$ supergravity}, Ph.D. thesis, Groningen University, 1990.
\bm{conformforPoin} M. Kaku and P.K. Townsend, {\sl Poincar\'e
    supergravity as broken superconformal gravity}, Phys. Lett. {\bf 76B}
    (1978) 54.

\bm{N1YMmsg} E. Cremmer, S. Ferrara, L. Girardello, B. Julia, P. van
     Nieuwenhuizen, and J. Scherk, {\sl Spontaneous symmetry breaking
     without cosmological constant}, Nucl. Phys. {B147}, 105 (1979);\\
     E. Cremmer, S. Ferrara, L. Girardello and A. Van Proeyen,
     {\sl Yang--Mills theories with local supersymmetry : Lagrangian,
     transformation laws and superhiggs effect}, Nucl. Phys. {\bf B212}
     (1983) 413.

\bm{dWLVP} B. de Wit, P. Lauwers and A. Van Proeyen, {\sl Lagrangians
     of $N=2$ Supergravity-matter Systems}, Nucl. Phys. {\bf B255} (1985)
     569.

\bm{ber2} E. Bergshoeff, M. de Roo and B. de Wit, {\sl Extended
     conformal supergravity}, Nucl. Phys. {\bf B182} (1981) 173.
\bm{review} B. de Wit, {\sl Conformal invariance in extended supergravity},
         in {\it Supergravity'81}, eds. S. Ferrara and J.G. Taylor,
          (Cambridge University Press, 1982) 267;\\
A.~Van Proeyen,
{\sl Superconformal Tensor Calculus in $N=1$ and $N=2$ Supergravity}, in
{\it Supersymmetry and supergravity 1983},
ed. B.~Milewski (World Scientific, Singapore 1983) 93.

\bm{C1}   R. Corrado, B. Florea and  R. McNees,
          {\sl Correlation functions of operators and Wilson surfaces
         in the d=6, (0,2) theory in the large N limit},
         {\tt hep-th/9902153}.

\bm{C2}   D. Berenstein, R. Corrado, W. Fischler and  J. Maldacena,
          {\sl The operator product expansion for Wilson loops and surfaces
          in the large N limit},
          {\tt hep-th/9809188}.



\bm{rozali} R.G. Leigh and M. Rozali, {\sl The large N limit of the $(2,0)$
   superconformal field theory}, Phys. Lett. {\bf B431} (1998) 311,
   {\tt hep-th/9803068}.

\bm{g} M. G\"{u}naydin, P. van Nieuwenhuizen and N.P. Warner, {\sl
    General construction of the unitary representations of anti de Sitter
    superalgebras and the spectrum of the $S^4$ compactifications of eleven
    dimensional supergravity}, Nucl. Phys. {\bf B255} (1985) 63.

\bibitem{conf98} J-H.~Park,
{\sl Superconformal Symmetry in Six-dimensions and Its Reduction to Four},
Nucl. Phys. {\bf B539}, 599 (1999), {\tt hep-th/9807186}.



\end{thebibliography}
\end{document}